\documentclass[prl,twocolumn,showpacs]{revtex4}
\usepackage{amsmath}
\usepackage{dcolumn}
\usepackage{graphicx}
\usepackage{bm}

\begin{document}

\title{String order and hidden topological symmetry in the $SO(2n+1)$
symmetric matrix product states}
\author{Hong-Hao Tu and Guang-Ming Zhang }
\email{gmzhang@tsinghua.edu.cn}
\affiliation{Department of Physics, Tsinghua University, Beijing 100084, China}
\author{Tao Xiang}
\affiliation{Institute of Physics, Chinese Academy of Sciences, P.O. Box 603, Beijing
100190, China; \\
Institute of Theoretical Physics, Chinese Academy of Sciences, P.O. Box
2735, Beijing 100190, China}
\date{\today}

\begin{abstract}
We have introduced a class of exactly soluble Hamiltonian with either $%
SO(2n+1)$ or $SU(2)$ symmetry, whose ground states are the $SO(2n+1)$
symmetric matrix product states. The hidden topological order in these
states can be fully identified and characterized by a set of nonlocal string
order parameters. The Hamiltonian possesses a hidden $(Z_{2}\times
Z_{2})^{n} $ topological symmetry. The breaking of this hidden symmetry
leads to $4^{n}$ degenerate ground states with disentangled edge states in
an open chain system. Such matrix product states can be regarded as cluster
states, applicable to measurement-based quantum computation.
\end{abstract}

\maketitle

Quantum spin systems have shown many fascinating phenomena and stimulated
great interest in the past decades. Based on semiclassical argument, Haldane
predicted that there is a finite excitation gap in the ground state of an
integer antiferromagnetic Heisenberg spin chain \cite{Haldane-1983}. This
intriguing feature of quantum spin chains results from the breaking of a
hidden topological symmetry embedded in the valence bond solid state
proposed by Affleck, Kennedy, Lieb, and Tasaki (AKLT) \cite{Affleck-1987}.
The valence bond solid is a matrix product state in one dimension. It shows
a striking analogy to the Laughlin ground state for the fractional quantum
Hall effect \cite{Arovas-1988,Girvin-1989}. To characterize this topological
symmetry, a set of nonlocal string order parameters were introduced \cite%
{den Nijs-1989,Kennedy-1991}. These string order parameters provide a
faithful quantification of the hidden antiferromagnetic order of the $S=1$
Heisenberg model. Associated with these order parameters, a nonlocal unitary
transformation can be constructed to expose explicitly the $Z_{2}\times
Z_{2} $ symmetry of the Hamiltonian \cite%
{Kennedy-1991,Oshikawa-1992,Suzuki-1995}. However, a nonlocal string order
parameter that reflects correctly the hidden $Z_{S+1}\times Z_{S+1}$
topological symmetry of the higher-$S$ valence bond solid has not been found
\cite{Tu-AKLT}.

In this paper, we introduce a novel matrix product state with $SO(2n+1)$
symmetry and show that it is the exact ground state of a model Hamiltonian
with nearest neighbor interactions constructed with either the $SO(2n+1)$
projection operators or more generally the $SU(2)$ spin projection
operators. Unlike the valence bond solid state, we find that the hidden
topological order in this class of matrix product states can be fully
identified and characterized by a set of nonlocal string order parameters.
When $n=1$, the $SO(3)$ symmetric matrix product state is exactly the same
as the $S=1$ valence bond solid state and the model Hamiltonian possesses a
hidden $Z_{2}\times Z_{2}$ topological symmetry \cite%
{Kennedy-1991,Oshikawa-1992,Suzuki-1995}. When $n>1$, it will be shown that
the $SO(2n+1)$ ground state possesses a hidden $(Z_{2}\times Z_{2})^{n}$
topological symmetry. The breaking of this hidden symmetry leads to $4^{n}$
degenerate ground states with disentangled edge states in an open chain
system.

Let us start by considering a one dimensional lattice system with $SO(2n+1)$
symmetry. Each lattice site contains $2n+1$ basis states $\{\left|
n^{a}\right\rangle ;$ $a=1,\cdots ,2n+1\}$, which can be rotated within the $%
SO(2n+1)$ space as follows
\begin{equation}
L^{ab}|n^{c}\rangle =i\delta _{bc}|n^{a}\rangle -i\delta _{ac}|n^{b}\rangle ,
\label{Rotation}
\end{equation}%
where $L^{ab}$ $(a<b)$ are the $(2n^{2}+n)$ generators of the $SO(2n+1)$ Lie
algebra, satisfying the following commutation relations
\begin{equation}
\lbrack L^{ab},L^{cd}]=i(\delta _{ad}L^{bc}+\delta _{bc}L^{ad}-\delta
_{ac}L^{bd}-\delta _{bd}L^{ac}).
\end{equation}%
According to the Lie algebra, the product of any two $SO(2n+1)$ vectors can
be decomposed as a sum of an $SO(2n+1)$ scalar $\underline{1}$, an
antisymmetric $SO(2n+1)$ tensor $\underline{2n^{2}+n}$, and a symmetric $%
SO(2n+1)$ tensor $\underline{2n^{2}+3n}$:
\begin{equation}
\underline{2n+1}\otimes \underline{2n+1}=\underline{1}\oplus \underline{%
2n^{2}+n}\oplus \underline{2n^{2}+3n}.  \label{eq:decom}
\end{equation}%
The number above each underline is the dimension of the irreducible
representation.

In the spinor representation, the $SO(2n+1)$ generators can be expressed as $%
\Gamma ^{ab}=[\Gamma ^{a},\Gamma ^{b}]/2i$, where $\Gamma ^{a}$ ($a=1\sim
2n+1$) are the $2^{n}\times 2^{n}$ matrices that satisfy the Clifford
algebra $\{\Gamma ^{a},\Gamma ^{b}\}=2\delta _{ab}$ \cite{Georgi-1999}. For
each lattice site $i$, if the following matrix state is introduced
\begin{equation*}
g_{i}=\sum_{a}\Gamma ^{a}\left\vert n^{a}\right\rangle _{i},
\end{equation*}%
then it can be readily shown that the bond product of $g_{i}$ at any two
neighboring sites have finite projection only in the scalar $\underline{1}$
and the antisymmetric $\underline{2n^{2}+n}$ subspaces spanned by $%
|n_{i}^{a}\rangle $ and $|n_{i+1}^{a}\rangle $ states, because the product
of $\Gamma ^{a}$ and $\Gamma ^{b}$ can be expressed as $\Gamma ^{a}\Gamma
^{b}=\delta _{ab}+i\Gamma ^{ab}$. This is a special property of the $SO(2n+1)
$ spinor representation constructed by Clifford algebra. By applying this
argument to a periodic chain, we can show that the matrix product state
defined by
\begin{equation}
|\Psi \rangle =\mathrm{Tr}\left( g_{1}g_{2}\ldots g_{L}\right) ,
\label{eq:MPS}
\end{equation}%
is the exact ground state of the following $SO(2n+1)$ symmetric Hamiltonian
\begin{equation}
H_{SO(2n+1)}=\sum_{i}\mathcal{P}_{\underline{2n^{2}+3n}}(i,i+1),
\label{eq:soham}
\end{equation}%
where $\mathcal{P}_{\underline{2n^{2}+3n}}(i,j)$ is a projection operator
that projects the states at sites $i$ and $j$ onto their $SO(2n+1)$
symmetric tensor $\underline{2n^{2}+3n}$. To compute the static correlation
functions of the matrix product ground state (\ref{eq:MPS}), we can use a
transfer matrix method \cite{Klumper-1991,Suzuki-1995}. At large distance,
the two-point correlation functions of $SO(2n+1)$ generators decay
exponentially as%
\begin{equation}
\left\langle L_{i}^{ab}L_{j}^{ab}\right\rangle \sim \exp \left( -\frac{|j-i|%
}{\xi }\right) ,  \label{eq:decay}
\end{equation}%
with the correlation length $\xi =1/\ln \left\vert \frac{2n+1}{2n-3}%
\right\vert $.

For the three $SO(2n+1)$ channels given in Eq. (\ref{eq:decom}), the bond
Casimir charge $\sum_{a<b}(L_{i}^{ab}+L_{j}^{ab})^{2}$ for two adjacent
sites takes the values $0$, $4n-2$, and $4n+2$, respectively. Combining this
result with the equation $\sum_{a<b}(L_{i}^{ab})^{2}=2n$ and the
completeness condition of the projection operators, we can then express the
bond projection operator $\mathcal{P}_{\underline{2n^{2}+3n}}(i,j)$ with the
$SO(2n+1)$ generators as
\begin{eqnarray*}
&&\mathcal{P}_{\underline{2n^{2}+3n}}(i,j) \\
&=&\frac{1}{2}\sum_{a<b}L_{i}^{ab}L_{j}^{ab}+\frac{1}{4n+2}%
(\sum_{a<b}L_{i}^{ab}L_{j}^{ab})^{2}+\frac{n}{2n+1}.
\end{eqnarray*}%
Thus the model defined by Eq. (\ref{eq:soham}) is a bilinear-biquadratic
Hamiltonian in terms of the $SO(2n+1)$ generators.

At each lattice site, the $2n+1$ vectors of $SO(2n+1)$ can be also
constructed from the $S=n$ quantum spin states. In the $SU(2)$ spin
language, the last two channels in Eq. (\ref{eq:decom}) correspond to the
total bond spin $S=1,3,\ldots ,2n-1$ and $S=2,4,\ldots ,2n$ states,
respectively. Furthermore, it can be shown that the bond projection
operators of $SO(2n+1)$ can be expressed using the spin projection operators
$P_{S=m}(i,j)$ as
\begin{eqnarray*}
\mathcal{P}_{\underline{2n^{2}+n}}(i,j) &=&\sum_{m=1}^{n}P_{S=2m-1}(i,j), \\
\mathcal{P}_{\underline{2n^{2}+3n}}(i,j) &=&\sum_{m=1}^{n}P_{S=2m}(i,j).
\end{eqnarray*}%
Thus $\mathcal{P}_{\underline{2n^{2}+3n}}(i,j)$ is to project the spin
states at sites $i$ and $j$ onto the nonzero even total spin states. Based
on this property, we can further show that the matrix product wavefunction (%
\ref{eq:MPS}) is also the ground state of the following integer spin
Hamiltonian
\begin{equation}
H_{SU(2)}=\sum_{i}\sum_{m=1}^{n}J_{m}P_{S=2m}(i,i+1)  \label{eq:model2}
\end{equation}%
with all $J_{m}>0$. This model is $SU(2)$-invariant in general. However, the
ground state (\ref{eq:MPS}) possesses an \textit{emergent} $SO(2n+1)$
symmetry. When all $J_{m}=1$, $H_{SU(2)}$ becomes $SO(2n+1)$-invariant. In
this case, $H_{SU(2)}$ simply reduces to $H_{SO(2n+1)}$.

It is interesting to compare $H_{SU(2)}$ with the AKLT model of valence bond
solid proposed by Affleck \textit{et. al}. \cite{Affleck-1987,Arovas-1988}
\begin{equation}
H_{\text{\textrm{AKLT}}}=\sum_{i}\sum_{m=n+1}^{2n}K_{m}P_{S=m}(i,i+1)
\end{equation}%
with all $K_{m}>0$. The ground state of $H_{\mathrm{AKLT}}$ is also a matrix
product state similar to Eq. (\ref{eq:MPS}), but $g_{i}$ is now a $%
(S+1)\times (S+1)=(n+1)\times (n+1)$ matrix \cite{Suzuki-1995}. These two
matrix product states have different topological properties and belong to
different topological phases when $n>1$. Therefore $H_{SU(2)}$ and $%
H_{SO(2n+1)}$ can be viewed as a new family of exactly solvable quantum
integer spin models to understand the internal\ structures of Haldane gap
phases.

When $n=1$, both $H_{SO(2n+1)}$ and $H_{SU(2)}$ become exactly the same as
the $S=1$ AKLT model $H_{\mathrm{AKLT}}$. The ground state has a hidden
antiferromagnetic order in which the up and down spins lie alternately along
the lattice, sandwiched by arbitrary number of non-polarized spin states.
This dilute antiferromagnetic order can be measured by a nonlocal string
order parameter first proposed by den Nijs and Rommelse \cite{den Nijs-1989}%
,
\begin{equation}
\mathcal{O}^{\mu }=\lim_{\left\vert j-i\right\vert \rightarrow \infty
}\langle S_{i}^{\mu }\prod_{l=i}^{j-1}e^{i\pi S_{l}^{\mu }}S_{j}^{\mu
}\rangle =\frac{4}{9},\,  \label{eq:string}
\end{equation}%
where $\mu =x$, $y$ or $z$. By performing a nonlocal unitary transformation
\cite{Kennedy-1991,Oshikawa-1992,Suzuki-1995} to the spin operators with the
following unitary operators
\begin{equation}
U=\prod_{j<i}\exp (i\pi S_{j}^{z}S_{i}^{x}),  \label{eq:KTtransf}
\end{equation}%
two of the above string order parameters are converted into the conventional
spin-spin correlation functions. The $SU(2)$ symmetry of the AKLT model is
then reduced to a discrete $Z_{2}\times Z_{2}$ symmetry \cite%
{Kennedy-1991,Oshikawa-1992,Suzuki-1995}. This reveals a hidden topological
symmetry of the original model. The breaking of this topological symmetry
leads to the opening of the Haldane gap and the four-fold degenerate ground
states in an open chain.

Similar to the $n=1$ case, the general $SO(2n+1)$ ($n>1$) matrix product
state (\ref{eq:MPS}) also contains interesting hidden antiferromagnetic
orders. Since $SO(2n+1)$ is a rank-$n$ algebra, one can always classify the
states at each site using $n$ quantum numbers (weights) $\{m_{1},\cdots
,m_{n}\}$ subjected to the constraint
\begin{equation}
m_{\alpha }m_{\beta }=0,\qquad (\alpha \not=\beta ).  \label{eq:constraint}
\end{equation}%
Here $\{m_{1},\cdots ,m_{n}\}$ are the eigenvalues of the mutually commuting
Cartan generators $\{L^{12},L^{34},\ldots ,L^{2n-1,2n}\}$
\begin{equation}
L^{2\alpha -1,2\alpha }|m_{\alpha }\rangle =m_{\alpha }|m_{\alpha }\rangle
,\quad (m_{\alpha }=0,\pm 1).
\end{equation}%
According to Eq. (\ref{Rotation}), all these Cartan generators annihilate
the state $\left\vert n^{2n+1}\right\rangle =\left\vert 0,0,\ldots
,0\right\rangle $ . The other basis states are given by
\begin{equation}
\left\vert 0\cdots ,m_{\alpha }=\pm 1,\cdots 0\right\rangle =\frac{1}{\sqrt{2%
}}\left( \left\vert n^{2\alpha }\right\rangle \pm i\left\vert n^{2\alpha
-1}\right\rangle \right) .
\end{equation}%
From the property of the Clifford algebra, the hidden antiferromagnetic
order of the ground state $|\Psi \rangle $ can now be identified. In any of
these $m_{\alpha }\,(\alpha =1\sim n)$ channel, it can be shown that $%
|m_{\alpha }\rangle $ is dilute antiferromagnetically ordered, same as for
the $S=1$ valence bond solid. Namely, the states of $m_{\alpha }=1$ and $%
m_{\alpha }=-1$ will alternate in space if all the $m_{\alpha }=0$ states
between them are ignored. For example, a typical configuration of the ground
state of the $SO(5)$ system is
\begin{equation*}
\begin{array}{crcccccccccccccccl}
m_{1}: & \quad \cdots & 0 & \uparrow & 0 & 0 & \downarrow & \uparrow & 0 & 0
& 0 & \downarrow & \uparrow & 0 & \downarrow & 0 & \uparrow & \cdots \\
m_{2}: & \cdots & \uparrow & 0 & \downarrow & 0 & 0 & 0 & \uparrow &
\downarrow & 0 & 0 & 0 & \uparrow & 0 & \downarrow & 0 & \cdots%
\end{array}%
\end{equation*}%
where $(\uparrow ,0,\downarrow )$ represent $|m\rangle =\left( |1\rangle
,|0\rangle ,\left\vert -1\right\rangle \right) $ states, respectively.

This hidden antiferromagnetic order reminds us a generalization of the den
Nijs-Rommelse nonlocal string order parameters to characterize this state.
Similar to Eq. (\ref{eq:string}) of the $n=1$ case \cite{den Nijs-1989}, the
string order parameters can be defined as
\begin{equation}
\mathcal{O}^{ab}=\lim_{|j-i|\rightarrow \infty }\langle
L_{i}^{ab}\prod_{l=i}^{j-1}\exp (i\pi L_{l}^{ab})L_{j}^{ab}\rangle .
\label{eq:SOP}
\end{equation}%
Since the ground state is $SO(2n+1)$ rotationally invariant, the above
nonlocal order parameters should all be equal to each other. Thus to
determine the value of these parameters, only the value of $\mathcal{O}^{12}$
needs to be evaluated. In the $L^{12}$ channel, the role of the phase factor
in Eq. (\ref{eq:SOP}) is to correlate the finite spin polarized states in
the $m_{1}$ channel at the two ends of the string. If nonzero $m_{1}$ takes
the same value at the two ends, then the phase factor is equal to $1$. On
the other hand, if nonzero $m_{1}$ takes two different values at the two
ends, then the phase factor is equal to $-1$. Thus the value of $\mathcal{O}%
^{12}$ is determined purely by the probability of $m_{1}=\pm 1$ appearing at
the two ends of the string. Since the ground state is translation invariant,
it is straightforward to show that the probability of the states $m_{1}=\pm
1 $ appearing at one lattice site is $2/(2n+1)$ and thus $\mathcal{O}%
^{12}=4/(2n+1)^{2}$.

\begin{figure}[tbp]
\includegraphics[scale=0.5]{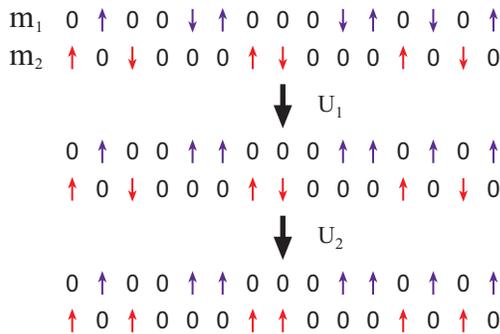}
\caption{(Color online) Changes of a typical configuration of the $SO(5)$
ground state under the unitary transformation defined by Eq. (\protect\ref%
{eq:transf}). $U_{1}$ and $U_{2}$ transform successively all $m_{1}$ and $%
m_{2}$ states to two diluted ferromagnetic configurations, respectively.}
\label{fig:transf}
\end{figure}

The Kennedy-Tasaki unitary transformations (\ref{eq:KTtransf}) for $n=1$
case \cite{Kennedy-1991,Oshikawa-1992,Suzuki-1995} can also be generalized
to arbitrary $n>1$ cases. In the $SO(2n+1)$ Lie algebra, $(L^{2\alpha
-1,2\alpha },L^{2\alpha -1,2n+1},L^{2\alpha ,2n+1})$ span an $SO(3)$
sub-algebra in which $\exp (i\pi L^{2\alpha ,2n+1})$ plays the role of
flipping the quantum number $m_{\alpha }$. This exponential operator can
flip the quantum numbers of $m_{\alpha }$ without disturbing the quantum
states in all other channels. This indicates that if we take the following
nonlocal unitary transformation in the $m_{\alpha }$ channel
\begin{equation}
U_{\alpha }=\prod_{j<i}\exp \left( i\pi L_{j}^{2\alpha -1,2\alpha
}L_{i}^{2\alpha ,2n+1}\right) ,
\end{equation}%
then all the configurations in this channel will be ferromagnetically
ordered. Furthermore, by performing this nonlocal transformation
successively in all the channels
\begin{equation}
U=\prod_{\alpha =1}^{n}U_{\alpha },  \label{eq:transf}
\end{equation}%
then all the configurations of the ground state will become
ferromagnetically ordered. As an example, Fig. \ref{fig:transf} shows how
the $SO(5)$ matrix product state $|\Psi \rangle $ is successively changed
under this nonlocal unitary transformation.

By applying the generalized unitary transformation (\ref{eq:transf}) to the
\textit{Cartan} generators, it can be shown that
\begin{equation}
UL_{i}^{ab}U^{-1}=L_{i}^{ab}\exp (i\pi \sum_{j=1}^{i-1}L_{j}^{ab}).
\end{equation}%
Substituting this formula to Eq. (\ref{eq:SOP}), we find that
\begin{equation}
\mathcal{O}^{ab}=\lim_{\left\vert j-i\right\vert \rightarrow \infty
}\left\langle L_{i}^{ab}L_{j}^{ab}\right\rangle _{U}.  \label{eq:CF}
\end{equation}%
Thus the nonlocal string order parameters $\mathcal{O}^{ab}$ become the
ordinary correlation functions of local operators after the unitary
transformation.

Under the above transformation, the symmetry of the original Hamiltonian $%
H_{SO(2n+1)}$ is reduced, and determined by the symmetry of the unitary
transformation operators. In the $m_{\alpha }$ channel, it can be shown that
the unitary operator $U_{\alpha }$ possesses only a $Z_{2}\times Z_{2}$
symmetry \cite{Kennedy-1991,Oshikawa-1992,Suzuki-1995}. Therefore, the
Hamiltonian after the transformation has a $(Z_{2}\times Z_{2})^{n}$
symmetry. This is the hidden topological symmetry of the original
Hamiltonian $H_{SO(2n+1)}$, associated with the hidden topological order of
the original matrix product state $|\Psi \rangle $. Furthermore, the unitary
transformation (\ref{eq:transf}) breaks the translational symmetry. When it
is applied to an open chain system, the hidden $(Z_{2}\times Z_{2})^{n}$
topological symmetry of the Hamiltonian will be further broken, yielding $%
2^{n}$ free edge states at each end of the chain. Therefore, the open chain
has totally $4^{n}$ degenerate ground states, which can be distinguished by
their edge states.

As already mentioned, $H_{SO(2n+1)}$ is a bilinear-biquadratic Hamiltonian
in terms of the $SO(2n+1)$ generators. Actually, we can introduce a general
one-parameter family of the $SO(2n+1)$ bilinear-biquadratic model as
\begin{equation}
H=\sum_{i}\left[ \cos \theta \sum_{a<b}L_{i}^{ab}L_{i+1}^{ab}+\sin \theta
\left( \sum_{a<b}L_{i}^{ab}L_{i+1}^{ab}\right) ^{2}\right] ,  \label{eq:BB}
\end{equation}%
which is an extension of the quantum spin-$1$ bilinear-biquadratic model. To
determine the region of the Haldane gapped phase, we need to identity
several special integrable points. At $\theta _{1}=\tan ^{-1}\frac{1}{2n-1}$%
, the model (\ref{eq:BB}) becomes the Uimin-Lai-Sutherland (ULS)\ model with
an enhanced $SU(2n+1)$ symmetry, which can be solved by Bethe ansatz \cite%
{Sutherland-1975}. It is well-known that this model has gapless excitations
described by $SU(2n+1)_1$ Wess-Zumino-Witten model \cite{Affleck-1986}.
Based on the renormalization group approach, for $\theta <\theta _{1}$, Itoi
and Kato \cite{Itoi-1997} found that the marginally relevant interaction
generates the Haldane gap, and the transition at the ULS point belongs to
the universality class of the Kosterlitz-Thouless phase transition.

One the other hand, using quantum inverse scattering methods, Reshetikhin
\cite{Reshetikhin-1983} had discovered another class of one-dimensional
quantum integrable $SO(n)$ model, corresponding to the point $\theta
_{2}=\tan ^{-1}\frac{2n-3}{(2n-1)^{2}}$ , where there are also gapless
excitations above the ground state. For $n=1$, this point corresponds to the
quantum spin-$1$ Takhatajan-Babujian model \cite{Takhatajan-1982}, which is
at the boundary between Haldane gap phase and dimerized phase. These
rigorous results suggest that the Haldane gapped phase for the general model
(\ref{eq:BB}) exists in the region
\begin{equation}
\tan ^{-1}\frac{2n-3}{(2n-1)^{2}}<\theta <\tan ^{-1}\frac{1}{2n-1}.
\end{equation}%
The exactly soluble point $\theta _{\text{MPS}}=\tan ^{-1}\frac{1}{2n+1}$
has been included. In the whole region, we expect that the system has an
energy gap in the excitations and the ordinary correlation functions display
exponentially decay. However, a nonvanishing string order parameter (\ref%
{eq:SOP}) can measure the breaking of the hidden topological symmetry.

For $n=1$, the spin-$1$ quantum antiferromagnetic Heisenberg model ($\theta
=0$) is just included in this region, however, we find that the $SO(2n+1)$
Heisenberg point for $n\geq 2$ does not belong to the Haldane gap phase. In
particular, when $n=2$, the corresponding $SO(5)$ antiferromagnetic
Heisenberg model has been used by Scalapino \textit{et. al}. \cite%
{Scalapino-1998} to describe the $SO(5)$ ``superspin''\ phase on a ladder
system of interacting electrons. Therefore, the ground state and low-lying
excitations of the quantum $SO(2n+1)$ symmetric generalized Heisenberg model
for $n\geq 2$ deserves further studies.

In conclusion, we have constructed an $SO(2n+1)$-invariant matrix product
state and shown that it is the exact ground state of an $SO(2n+1)$-symmetric
Hamiltonian defined by Eq. (\ref{eq:soham}) or more generally an $SU(2)$%
-symmetric spin Hamiltonian defined by Eq. (\ref{eq:model2}). This matrix
product state contains diluted antiferromagnetic orders in $n$ different
channels and a hidden $(Z_{2}\times Z_{2})^{n}$ topological symmetry. These
topological long range order can be characterized by a set of nonlocal
string order parameters. The breaking of the $(Z_{2}\times Z_{2})^{n}$
topological symmetry leads to the opening of an excitation gap between the
ground state and the first excitation state. In an open chain system, the $%
4^{n}$ edge states become completely disentangled and the ground states are $%
4^{n}$ degenerate. The multiple $Z_{2}$ nature of these topological states
suggests that they can serve as a resource of multiple qubits. We believe
that these states, similar as for the $S=1$ AKLT valence bond state, can be
encoded to perform ideal quantum teleportation \cite{Verstraete-Cirac-2004}
or fault-tolerant quantum computation through local spin measurements.

We acknowledge the support of NSF-China and the National Program for Basic
Research of MOST, China.

\end{document}